\documentclass[aps,twocolumn,pra,superscriptaddress,amsmath,showpacs,tightenlines]{revtex4}
\usepackage{epsfig,graphicx,times}
\usepackage{amstext}
\usepackage{amsmath}            
\usepackage{amssymb}            
\usepackage{graphicx}           
\usepackage{latexsym}
\usepackage{bm}
\usepackage{color}

\def \naturephys{{Nat. Phys.}}
\def \nature{{Nature (London)}}

\begin{document}

\title{From blockade to transparency: controllable photon transmission through a circuit QED system}

\author{Yu-xi Liu}\email{yuxiliu@mail.tsinghua.edu.cn}
\affiliation{Institute of Microelectronics, Tsinghua University,
Beijing 100084, China} \affiliation{Tsinghua National Laboratory
for Information Science and Technology (TNList), Beijing 100084,
China}\affiliation{CEMS, RIKEN, Saitama 351-0198, Japan}

\author{Xun-Wei Xu}
\affiliation{Institute of Microelectronics, Tsinghua University,
Beijing 100084, China}

\author{Adam Miranowicz}
\affiliation{Faculty of Physics, Adam Mickiewicz University,
61-614 Pozna\'n, Poland} \affiliation{CEMS, RIKEN, Saitama
351-0198, Japan}

\author{Franco Nori}
\affiliation{CEMS, RIKEN, Saitama 351-0198,
Japan}\affiliation{Physics Department, The University of Michigan,
Ann Arbor, Michigan 48109-1040, USA}
\date{\today}

\begin{abstract}
A strong photon-photon nonlinear interaction is a necessary
condition for photon blockade. Moreover, this nonlinearity can
also result a bistable behavior in the cavity field. We analyze
the relation between detecting field and photon blockade in a
superconducting circuit QED system, and show that photon blockade
cannot occur when the detecting field is in the bistable regime.
This photon blockade is the microwave-photonics analog of the
Coulomb blockade. We further demonstrate that the photon
transmission through such system can be controlled (from photon
blockade to transparency) by the detecting field. Numerical
calculations show that our proposal is experimentally realizable
with current technology.


\end{abstract}

\maketitle \pagenumbering{arabic}

\section{Introduction}

Superconducting circuit quantum electrodynamics (QED) allows
studying the interaction between superconducting qubits (or
superconducting artificial atoms) and quantized microwave fields
(see, e.g., the reviews~\cite{You11,Buluta11,Schoelkopf08}). The
coupling strengths have been explored from the strong-coupling
regime to the ultrastrong
one~\cite{Bourassa09,Niemczyk10,Forn10,Ashhab10}. It is well known
that the equally-spaced energy structure of the quantized
microwave field can be changed to an anharmonic one, including
dressed states~\cite{Liu06,Wilson07,Wilson10}. The nonlinear
energy splitting in circuit QED has been experimentally shown by
the Rabi frequencies for different numbers of microwave quanta
inside the cavity~\cite{Hofheinz08}. Moreover, the experimental
spectrum with the square root of the photon-number nonlinearity
was reported~\cite{Fink08}. Furthermore, the nonlinear response of
the vacuum Rabi resonance was demonstrated~\cite{Bishop09}.

Current experimental data indicate that the nonlinearity of the
microwave photons can be many orders of magnitude larger than that
of macroscopic media. These
experiments~\cite{Wilson07,Wilson10,Hofheinz08,Fink08} lay a solid
foundation for developing microwave nonlinear interactions, which
might be used to improve qubit
readout~\cite{Ginossar10,Boissonneault10}, and open the door to
further study microwave nonlinear quantum optics at the level of
single artificial atoms and single microwave photons. For example,
the photon blockade
phenomenon~\cite{Imamoglu97,Birnbaum05,Faraon08,Miranowicz13},
where subsequent photons are prevented from resonantly entering a
cavity, has recently been observed in circuit QED systems with
resonant~\cite{Lang11} and dispersive~\cite{Hoffman11}
qubit-cavity-field interactions.

\begin{figure}
\includegraphics[bb=14 220 590 460, width=8.5 cm, clip]{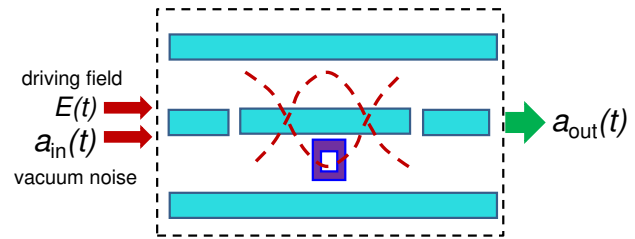}
\caption[]{Schematic diagram for the circuit QED system of a
superconducting charge qubit (denoted by the small purple square)
coupled to a transmission line resonator, indicated by the black
dashed box. The input (output) is denoted by the arrows going to
(the arrow leaving from) the black dashed box. We assume that the
input field, including the classical driving field $E(t)$ and the
vacuum noise $a_{\rm in}(t)$, is applied to the cavity at the left
port. The output field is measured at the right
port.}\label{fig1-1}
\end{figure}

Photon blockade originates from the anharmonic energy-level
structure of the light field when the strong photon-photon
interaction is induced by the nonlinear
medium~\cite{Imamoglu97,Miranowicz13}. In circuit QED systems in
resonance, the anharmonicity of the microwave field is from a
highly hybridization of the qubit and the microwave cavity
field~\cite{Lang11}. However, for the nonresonant case, the qubit
can induce the photon-photon interaction when the qubit degrees of
freedom are adiabatically eliminated~\cite{Hoffman11}. This photon
blockade is the analogue of Coulomb blockade~\cite{Fulton87},
where single-electron transport, through a small metallic or
semiconductor island sandwiched by two tunnel junctions in
electron devices, occurs one by one due to the Coulomb
interaction. Therefore, similar to the single-electron devices
using the Coulomb blockade, the photon blockade could be used as a
single-photon source or single-photon transistor.

One of the basic conditions for photon blockade is that the decay
rate of the cavity field should be less than the photon-photon
interacting strength. However, the decay of the photon-number
states is very different from those of coherent
states~\cite{Lu89,Wang08}; thus, photon blockade might not be
observed when the photon number inside the cavity is extremely
large. In addition, bistability is one of the basic properties of
nonlinear systems, but there is a lack of studies on the effect of
bistability on the photon blockade. Here we will study the
relation between bistability and photon blockade and explore the
possibility of unblocking the photon of the detecting field by
using electromagnetically-induced transparency~\cite{Liu12,Ian10}.

In our paper, we will first describe the model Hamiltonian in
Sec.~II, and also give detailed comparison between the photon and
the Coulomb blockades. Then in Sec.~III, we will study the
relation between the bistability and the photon blockade. In
Sec.~IV, we will discuss the effect of the driving field on the
photon blockade. In Sec.~V, we will study how the photon blockade
can be lifted and thus the system would become transparent.
Finally, we summarize our results in Sec.~VI.

\begin{figure}
\includegraphics[width=8.1 cm, clip]{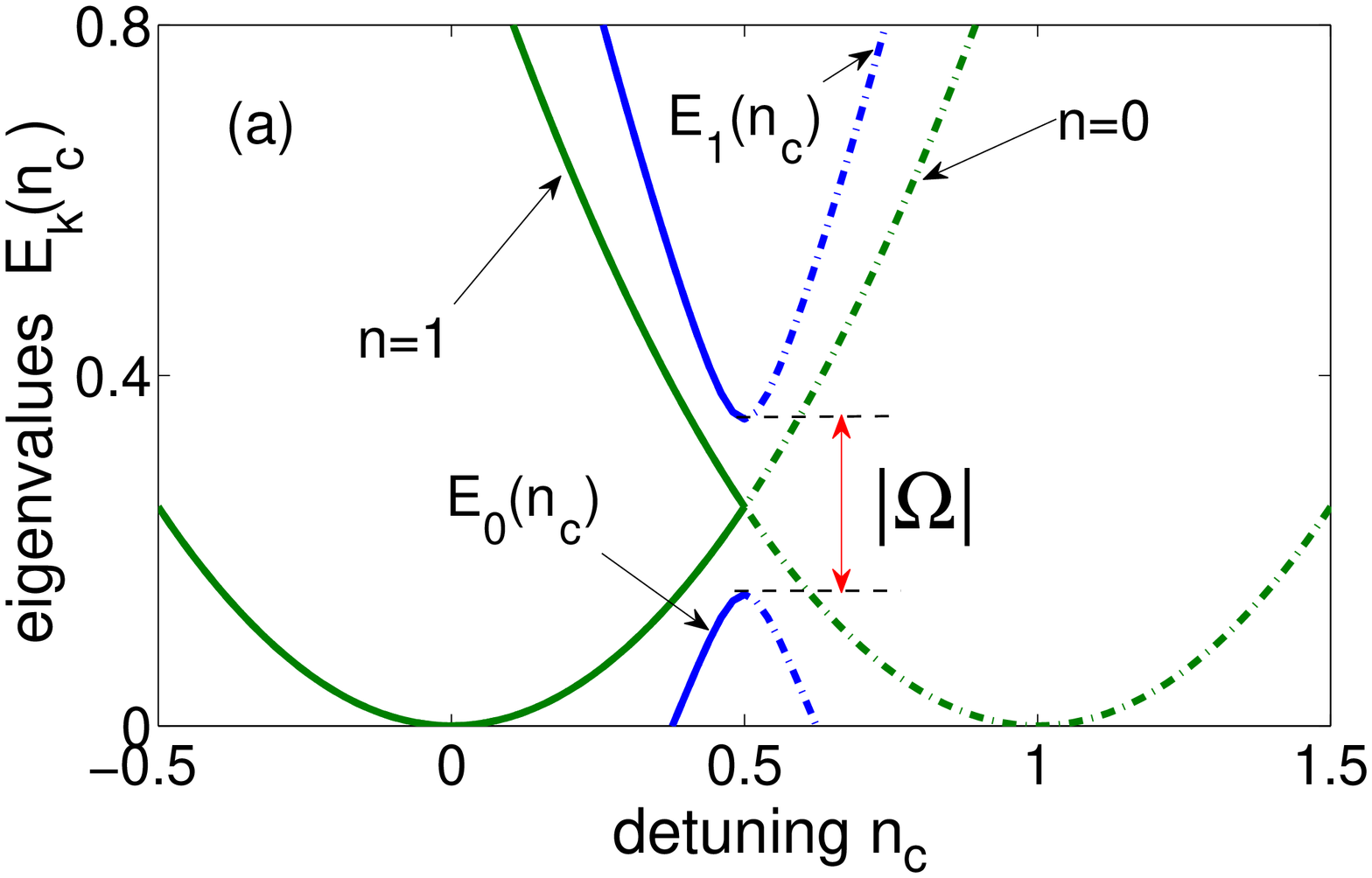}
\includegraphics[width=8 cm, clip]{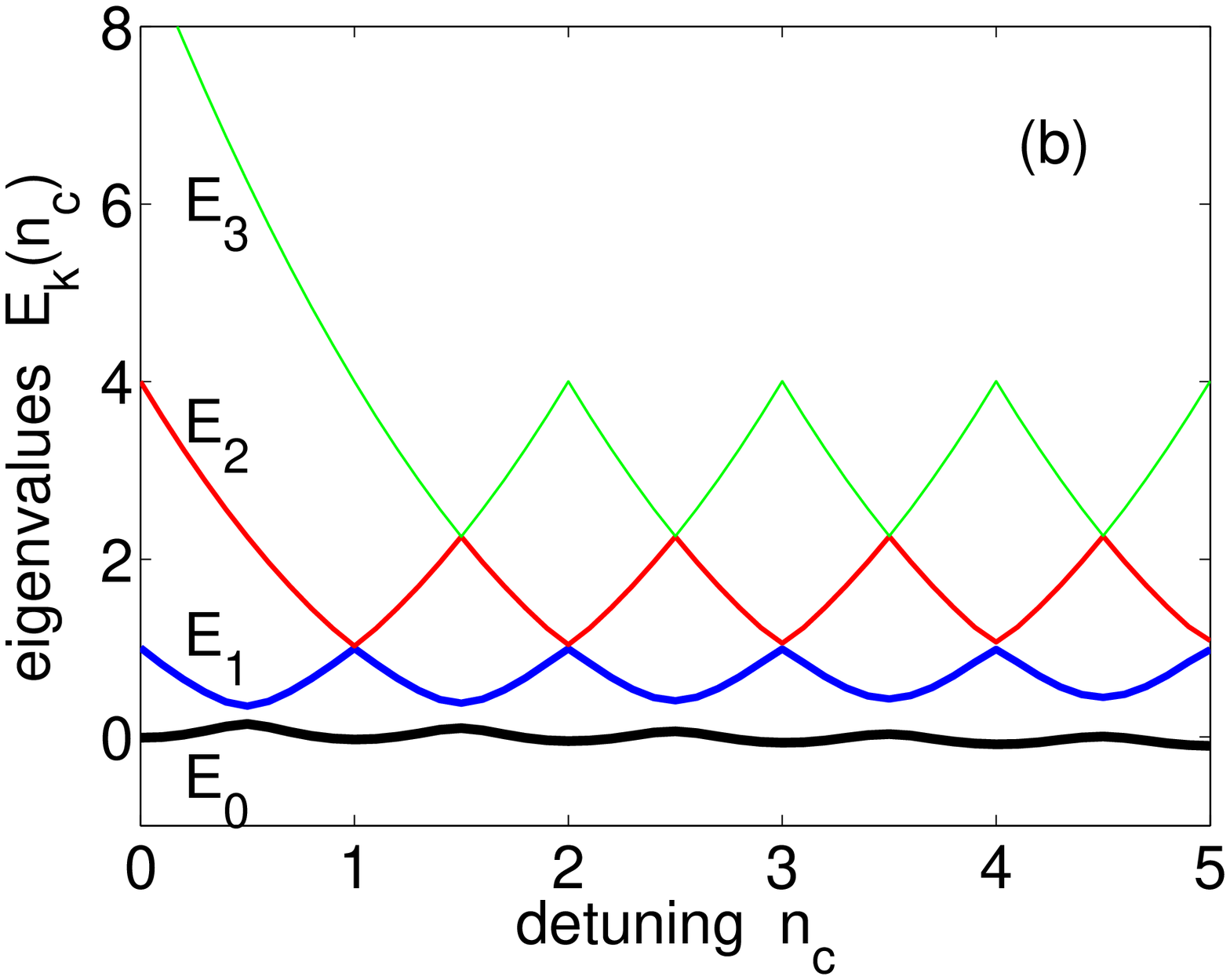}
\includegraphics[width=8 cm, clip]{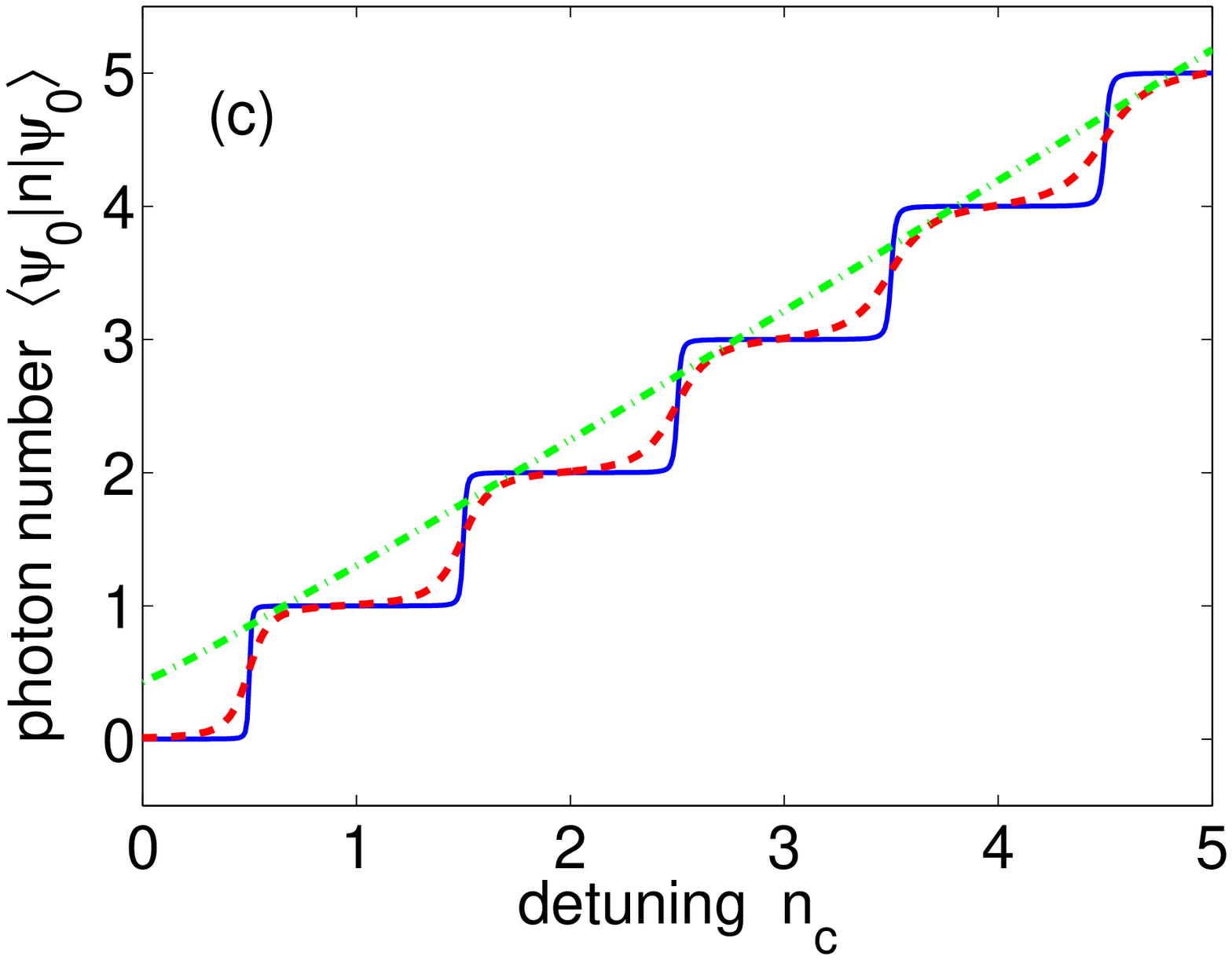}
\caption[]{(Color online) (a) Eigenvalues $E_k(n_{c})$ (green curves) of
the free Hamiltonian $\hbar \chi(n-n_{c})$ in Eq.~(\ref{eq:2})
versus the rescaled detuning $n_{c}=(\chi-\Delta_{d})/(2\chi)$ for
the photon numbers $n=0$ and $n=1$ inside the cavity, in energy
units of $\hbar \chi$. Blue curves denote the eigenvalues $E_{0}(n_{c})$ and $E_{1}(n_{c})$
of the ground and first excited states for the Hamiltonian in Eq.~(\ref{eq:2}) versus the rescaled detuning $n_{c}$ near
the point $n_{c}=0.5$, with an effective Hamiltonian $\hbar\chi(n_{c}-0.5)(|1\rangle\langle 1|-|0\rangle\langle 0|)
+\hbar(\Omega |1\rangle\langle 0|+\Omega^{*} |0\rangle\langle 1|)$, when the external field is included. The degeneracy
of the two eigenvalues in the free Hamiltonian is lifted at the
point $n_{c}=0.5$, with distance $\hbar|\Omega|$, by the external
field. All dot-dashed curves mean $n_{c}>0.5$, however solid curves mean $n_{c}<0.5$. (b) Several eigenvalues $E_{k}(n_{c})$  (up to four) of the Hamiltonian in Eq.~(\ref{eq:2}) versus the detuning $n_{c}$ for
the ratio $\chi/|\Omega|=10$. (c) The mean photon number in the
ground state $|\psi_{0}\rangle$ of the Hamiltonian in
Eq.~(\ref{eq:2}) versus the detuning $n_{c}$ for $\chi/|\Omega|=1$
(green dash-dotted curve), $10$ (red dashed curve), and $100$
(blue solid curve). \emph{This is the photon analog of the Coulomb
staircase. }}\label{fig1}
\end{figure}

\section{Model Hamiltonian}

As schematically shown in the black dashed box of
Fig.~\ref{fig1-1}, we study a circuit QED system in which the
microwave cavity field with frequency $\omega_{0}$ inside the
transmission line resonator, and the superconducting charge qubit
with frequency $\omega_{q}$ are coupled to each other with the
coupling strength $g$. We make the following assumptions:

(i) The cavity field and the qubit satisfy the large-detuning
condition $|\omega_{0}-\omega_{q}| \gg g$. That is, they are in
the dispersive interaction regime. Without loss of generality,
hereafter we also assume $\Delta\equiv \omega_{0}-\omega_{q} >0$.

(ii) The rotating wave approximation can be used. In this case, the dynamics of the interaction
Hamiltonian between the cavity field and the qubit can be
easily solved.

(iii) Other upper levels of the qubit system are far from the
first excited energy level, and the transition frequency between
the first and second excited states is much bigger than the
frequency of the cavity field. Thus, the cavity field only
interacts with the qubit.

(iv) The circuit QED system is in the bad-cavity-limit, i.e., the decay rate of the microwave cavity field is much higher than that of the qubit.

The above conditions can be satisfied in circuit QED
systems~\cite{You11,Schoelkopf08} by well chosen sample design and fabrication. The interaction between the
cavity field and the external environment, including the classical
driving field and the vacuum noise, can be described via the
input-output theory~\cite{Walls94}.

Let us first neglect the vacuum noise and assume that a weak
driving field $\varepsilon(t)$ is applied to the cavity; then the
driven Hamiltonian
\begin{equation}
\widetilde{H}=\hbar\omega_{0}a^{\dagger}a
+\hbar\frac{\omega_{q}}{2}\sigma_{z} +\hbar
g(a^{\dagger}\sigma_{-}+a\sigma_{+})
+\hbar[\varepsilon(t)a^{\dagger} +\text{H.c.}],
\end{equation}
under the rotating wave approximation, can be transformed to an
effective Hamiltonian $H=T^{\dagger}\widetilde{H} T$ with
\begin{eqnarray}\label{eq:1}
H=\hbar \omega_{0} a^{\dagger}a +
\frac{\hbar}{2}\left[\omega_{0}-E(N)\right]\sigma_{z}
+\hbar\left[\varepsilon (t) a^{\dagger}+ \varepsilon^{*}(t)
a\right]
\end{eqnarray}
in the dispersive qubit-cavity-field interaction regime by
applying a canonical transformation~\cite{Compagno80}
\begin{equation}
T=\exp\left[-\frac{\theta}{\sqrt{4N}}(a\sigma_{+}-a^{\dagger}\sigma_{-})\right],
\end{equation}
with $\tan\theta=-2g\sqrt{N}/\Delta$ and
\begin{equation}
E(N)=\sqrt{\Delta^2+4g^2 N}.
\end{equation}
Here the total number operator $N$ of excitations of the qubit and
the cavity field is given by
\begin{equation}
N=a^{\dagger}a+\frac{1}{2}(\sigma_{z}+1).
\end{equation}
$a^{\dagger}$and $a$ are the creation and annihilation operators
of the cavity field, respectively; $\sigma_z$ is Pauli's operator,
and $\sigma_+$ ($\sigma_-$) is the qubit raising (lowering)
operator. In the derivation of Eq.~(\ref{eq:1}), the terms
proportion to $\textit{O}(N^{-1/2})$ are neglected.

The excited and ground states of the qubit with sign $\pm $ for
$\sigma_{z}$ in the effective Hamiltonian in Eq.~(\ref{eq:1}) are
considered as the attractive and repulsive photon-photon
interactions for $\Delta>0$. However, they are considered as the
repulsive and attractive photon-photon interactions for
$\Delta<0$.

We also assume that the cavity field and the qubit are in the
strong-dispersive bad-cavity regime  as in
Refs.~\cite{Hoffman11,Bishop10}. That is, the decay rate $\kappa$
of the microwave cavity field is much higher than the decay
$\gamma$ and dephasing $\gamma_{\phi}$ rates of the qubit, and
also satisfies the condition,
\begin{equation}
\gamma, \gamma_{\phi} \ll \kappa \ll \frac{g^2}{\Delta}\ll g \ll
\Delta.
\end{equation}
Thus, the environmental effect on the qubit can be neglected in
our following discussions.

\begingroup \squeezetable
\begin{table}
\caption{\label{tab1} Equivalence between photon and Coulomb
blockades (or similarity between photon blockade and
superconducting charge qubits).}
\begin{ruledtabular}
\begin{tabular}{l|p{3.5cm} |p{4cm}}
Quanta & Photons & Electrons \\
\hline
Characteristic energy &Kerr energy $\hbar \chi$ & Charging energy $E_{c}$ \\
\hline
Energy offset control & Driving field frequency $\omega_{d}$& Gate voltage $V_{g}$\\
\hline
Detecting method & Driving field $E(t)$ & Bias voltage $V_{b}$\\
\hline
Measurement of output & Photon number $\langle n\rangle$ & Electron current $I$\\
\hline Tunneling energy &Coupling energy $\hbar|\Omega|$ &
Josephson energy $E_{J}$
\end{tabular}
\end{ruledtabular}
\end{table}
\endgroup

If a monochromatic driving field with frequency $\omega_{d}$ is
applied to the cavity mode; then the Hamiltonian in
Eq.~(\ref{eq:1}) with $\varepsilon(t)=\Omega e^{-i\omega_{d}t}$
can be used to describe the photon
blockade~\cite{Imamoglu97,Liu12} when the coupling strength
$\Omega$ between the driving field and the cavity field is much
smaller than the photon-photon coupling constant $\chi$ given
below. Moreover, the decay rate $\kappa$ of the cavity field
should also be much smaller than $\chi$ to experimentally observe
photon blockade. In the dispersive regime, the single two-level
atom-induced photon blockade can be understood by expanding the
Hamiltonian in Eq.~(\ref{eq:1}), up to third order in the
parameter $g/\Delta$, as
\begin{eqnarray}\label{eq:2}
H_{\rm eff}&=&\hbar \chi(n-n_{\rm c})^2+\hbar( \Omega
a^{\dagger}+\Omega^{*} a)
\end{eqnarray}
with $n=a^{\dagger}a$ in the rotating reference frame with the
frequency $\omega_{d}$ for the driving field. Here,
$\chi=g^4/\Delta^3$ denotes the photon-photon interaction
strength, $n_{c}=(\chi-\Delta_{d})/(2\chi)$ is the rescaled
detuning $\Delta_{d}=\omega_{0}-\omega_{d}$ between the driving
and the cavity fields. We notice that an effective Hamiltonian as
in Eq.~(\ref{eq:2}) can also be derived for the case that the
cavity field interacts with the multi-level superconducting
quantum systems. The detailed derivations are in
Ref.~\cite{Boissonneault10}.

In the derivation of Eq.~(\ref{eq:2}), we have used the
weak-excitation condition $\Delta > 2g\langle N\rangle$ for the
photon blockade, also several constant terms and the qubit
state-dependent cavity-frequency shift have been neglected with
the assumption $\sigma_{z}=1$. That is, the qubit is in its
excited state. Actually, the sign of the effective Hamiltonian in
Eq.~(\ref{eq:2}), derived from the Hamiltonian in
Eq.~(\ref{eq:1}), depends both on the qubit state and the detuning
$\Delta$ between the qubit frequency $\omega_{q}$ and the
cavity-field frequency $\omega_{q}$. In the following numerical
discussions, we use our assumption $\Delta=\omega_{0}-\omega_{q} >
0$ and $\sigma_{z}=1$ (the qubit is in the excited state). These
selections are only used for the numerical calculations. These
assumptions are equivalent to the case $\Delta<0$ and
$\sigma_{z}=-1$ (when the qubit is in the ground state).

Equation~(\ref{eq:2}) shows that the Kerr energy $\hbar \chi$
corresponds to the charging energy $E_c$, while the driving field
frequency $\omega_{d}$ (or, equivalently, the rescaled detuning
$n_{\rm c}$) corresponds to the gate voltage $V_g$ in
single-electron devices for Coulomb blockade (see Table~I).
Equation~(\ref{eq:2}) also shows that the parameters $\chi$,
$\omega_{d}$ (or $n_{\rm c}$), and $|\Omega|$ play similar
functions as the charging energy, the gate voltage, and the
Josephson energy of the charge qubit, respectively. For resonant
driving, $\Delta_{d}=0$ and $n_{\rm c}=0.5$, the photon states
$|0\rangle$ and $|1\rangle$ are degenerate for the free
Hamiltonian $\hbar \chi(n-0.5)^2$. As shown in Fig.~\ref{fig1}(a),
this degeneracy can be lifted by the coupling strength $|\Omega|$.
Figures~\ref{fig1}(b,c) show the variations of the eigenenergies
and the mean photon number $\langle n\rangle$ of the ground state
for the Hamiltonian in Eq.~(\ref{eq:2}) as functions of the
parameter $n_{\rm c}$. Variations for both the eigenenergy and the
mean photon number are the same as those of the eigenenergy and
mean charge number of charge qubits or single-electron devices for
Coulomb blockade. The staircase shape for the mean photon number
was experimentally demonstrated in circuit QED
system~\cite{Hoffman11}. Moreover, Fig.~\ref{fig1}(a) also shows
that a large ratio $\chi/|\Omega|$ corresponds to a sharper step.
Namely, a weak driving field and strong photon-photon interaction
are more useful for detecting photon blockade.

\begin{figure}
\includegraphics[bb=-1 -1 421 299, width=8.5 cm, clip]{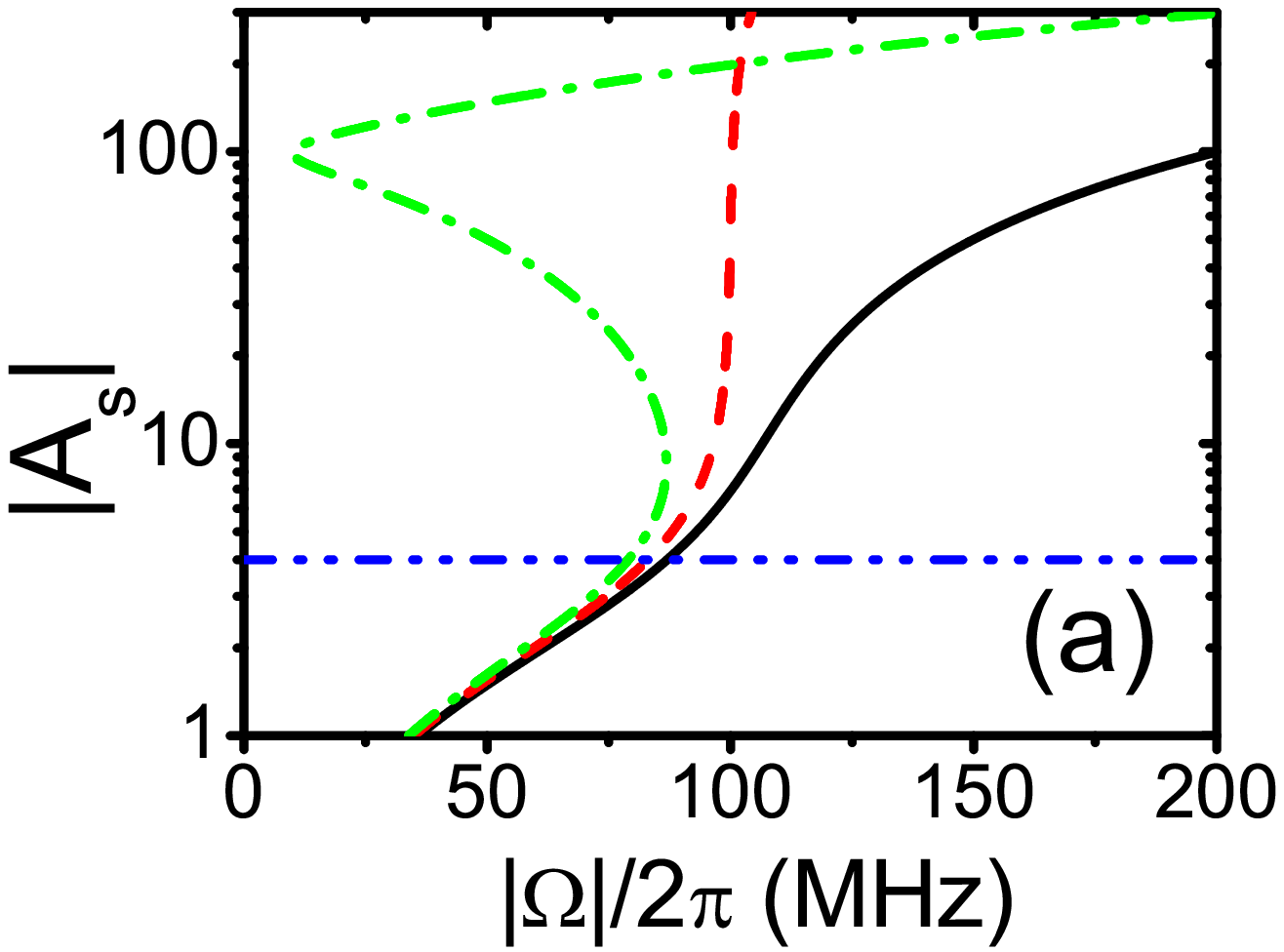}
\includegraphics[bb=-1 -1 421 299, width=8.5 cm, clip]{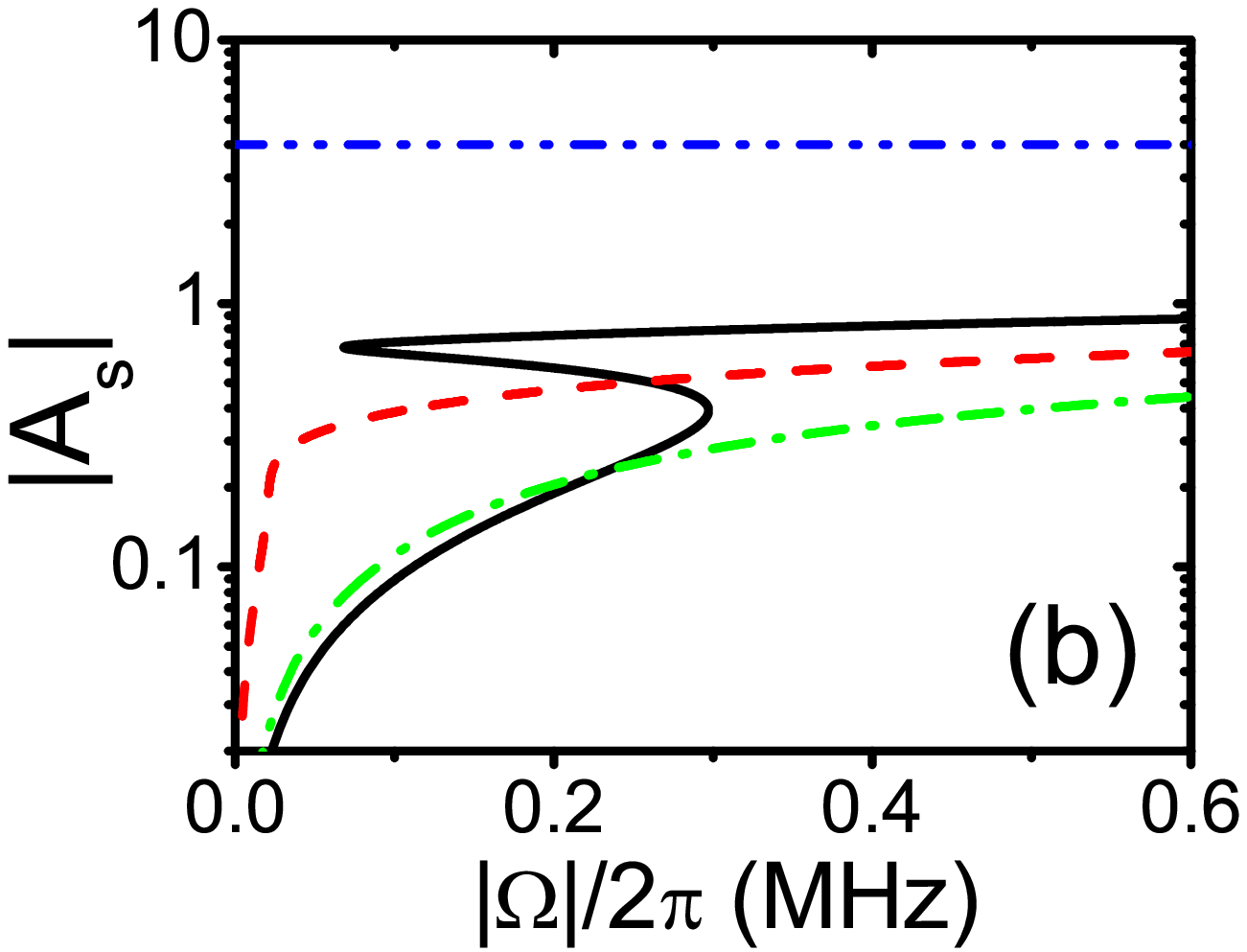}
\caption[]{(Color online) Logarithm ${\rm log}(|A_{s}|)$ of the
steady-state solution $|A_{s}|$ versus the strength of the driving
field $|\Omega|$, for different detunings $\Delta_{d}/2\pi$ equal
to: (a) $-1$ MHz (solid black), 0 MHz (dashed red) and  1 MHz
(dot-dashed green), and (b) 36 MHz (solid black), 37 MHz (dashed
red) and  38 MHz (dot-dashed green curves). Here, $N_{\rm up}$
(double-dot-dashed blue lines) is the upper bound photon number
for the photon blockade, $\Delta/2\pi=1$ GHz, $\kappa/2\pi=0.1$
MHz, and $g/2\pi=200$ MHz. For the above-given parameters,
$|A_{s}|$ has two stable states when $9.1\, {\rm kHz} <
\Delta_{\rm d}< 36.95\, {\rm MHz}$}\label{fig2}
\end{figure}

\section{Bistability and blockade}

We know that driven nonlinear photonic systems can exhibit
bistability. To study the relation between  bistability and photon
blockade, we now write the equation of motion, using the driven
Hamiltonian in Eq.~(\ref{eq:1}) with $\varepsilon(t)=\Omega
e^{-i\omega_{d}t}$, as
\begin{eqnarray}\label{eq:3}
\frac{\partial a}{\partial t}=-\left[i\omega_{0}+ \kappa
-i\frac{g^2\sigma_{z}}{E(N)}\right]a-i\Omega e^{-i\omega_{d}
t}-\sqrt{2\kappa}a_{\rm in}(t)
\end{eqnarray}
by using the relation $[a,f(a,a^{\dagger})]=\partial
f(a,a^{\dagger})/\partial a^{\dagger}$ between the operator $a$
and the function $f(a,a^{\dagger})$ of the operators $a$ and
$a^{\dagger}$. For example, we have the following relation
\begin{equation}
\left[a,\sqrt{\Delta^2+4g^2N}\right]=\frac{2g^2a}{\sqrt{\Delta^2+4g^2N}}.
\end{equation}
Here, we have neglected the detailed derivation of the decay rate
$\kappa$ and the quantum fluctuation $a_{\rm in}$ of the cavity
field and phenomenologically added them to Eq.~(\ref{eq:3})
according to Ref.~\cite{Walls94}. As schematically shown in
Fig.~\ref{fig1-1}, we notice that the input includes both the
quantum fluctuations $a_{\rm in}$ and the driving field $E(t)$. We
assume that the quantum fluctuations $a_{\rm in}$ due to the
vacuum field  are Gaussian and have a zero mean value $\langle
a_{\rm in}(t)\rangle =0 $ and satisfy the Markov correlation
\begin{equation}
\langle a_{\rm in}(t)a^{\dagger}_{\rm
in}(t^{\prime})\rangle=\delta(t-t^{\prime}).
\end{equation}
If we denote $a(t)=A(t)e^{-i\omega_{d} t}$,  then the steady-state
solution $A_{s} \equiv \langle A\rangle_{s}$ for the cavity field
becomes
\begin{equation}\label{eq:4}
A_{s}=-i\frac{\Omega}{\kappa+i[\Delta_{d}-g^2\sigma_{z}/E(\overline{N})]},
\end{equation}
by setting $(\partial a/\partial t)=0$ in Eq.~(\ref{eq:3}) and
using the mean-field approximation, e.g., $\langle
a^{\dagger}a\rangle_{s}= \langle a^{\dagger}\rangle_{s}\langle
a\rangle_{s}$.  We note in Eq.~(\ref{eq:4}) and hereafter,
although the operator $\sigma_{z}$ still remains in many
equations, the operator $\sigma_{z}$ is actually set to
$\sigma_{z}\equiv 1$ in all numerical calculations by assuming
that the qubit is always in its excited state. In this case, the
qubit, which is in the ground state, can also be easily discussed
by setting $\sigma_{z}\equiv-1$. We also note that
$E(\overline{N})$ in Eq.~(\ref{eq:4}) is the expression of $E(N)$
when replacing the operator $a^{\dagger}a=N$ by the steady-state
value $\overline{N}=|A_{s}|^2$, i.e.,
\begin{equation}\label{eq:4-1}
E(\overline{N})=\sqrt{\Delta^2+4g^2\left[|A_{s}|^2+\frac{1}{2}(1+\sigma_{z})\right]}.
\end{equation}
Equation~(\ref{eq:4}) shows that the cavity-field amplitude
$|A_{s}|^2$ has two stable states when $\Delta_{d}$ satisfies the
condition
\begin{equation}
\Delta_{-}<\Delta_{d}<\Delta_{+},
\end{equation}
with
\begin{equation}
\Delta_{\pm}=\frac{g^2}{E(\overline{N})}-\frac{2g^4|A_{s}|^2 \mp
\sqrt{4g^8|A_{s}|^4-\kappa^2
E^{6}(\overline{N})}}{E^{3}(\overline{N})}.
\end{equation}
Equation~(\ref{eq:4}) also clearly shows when the qubit is
decoupled from the cavity field (i.e., $g=0$) or when the driving
field makes $|A_{s}|^2$ extremely large, such that
$g\sigma_{z}/E(\overline{N})\approx 0$, then the response of the
system is the same as that of the harmonic oscillator. However,
when $g \neq 0$, the resonant peak of $|A_{s}|^2$ will move to
$\omega_{d}=\omega_{0}-g\sigma_{z}/E(\overline{N})$. That is, when
the cavity contains $|A_{s}|^2$ photons, the frequency
$\omega_{d}$ of the driving field should be increased an amount
$-g\sigma_{z}/E(\overline{N})$ to overcome the photon
blockade~\cite{Imamoglu97}. The upper-bound photon number inside
the cavity for the photon blockade is
\begin{equation}
N_{\rm up}
\;\sim\;\frac{g^4}{\Delta^3\kappa}\;=\;\frac{\chi}{\kappa}
\end{equation}
in the dispersive regime, as discussed in Ref.~\cite{Bishop10}.

In Fig.~\ref{fig2}, the steady-state $|A_{s}|$ versus the input
$|\Omega|$ is plotted for several different detunings $\Delta_{d}$
and other experimentally accessible parameters~\cite{Hoffman11}.
Figure~\ref{fig2} clearly shows that the bistability disappears
for either $\Delta_{d}>\Delta_{+}$ or $\Delta_{d}<\Delta_{-}$.
Figure~\ref{fig2} also shows that most values for $|A_{s}|$ are
smaller than the upper bound value $\sqrt{N_{\rm up}}$ for some
values of $\Delta_{d}$ in the bistable regime, e.g., values in the
region near $\Delta_{d}/2\pi=37$ MHz, as shown in
Fig.~\ref{fig2}(b). However, as shown in Fig.~\ref{fig2} (a) for
other parameter regimes of $\Delta_{d}$, we find that the upper
bound value $\sqrt{N_{\rm up}}$ can be smaller than some values of
$|A_{s}|$ corresponding to the lower branch of the bistable curve.
Therefore, Fig.~\ref{fig2} tells us that $N_{\rm up}$ is a
necessary but not sufficient condition for photon blockade.
Because one input corresponds to two stable outputs in the
hysteresis region, thus the photon blockade is not well defined in
such region.

\begin{figure}
\epsfxsize=8cm\epsfbox{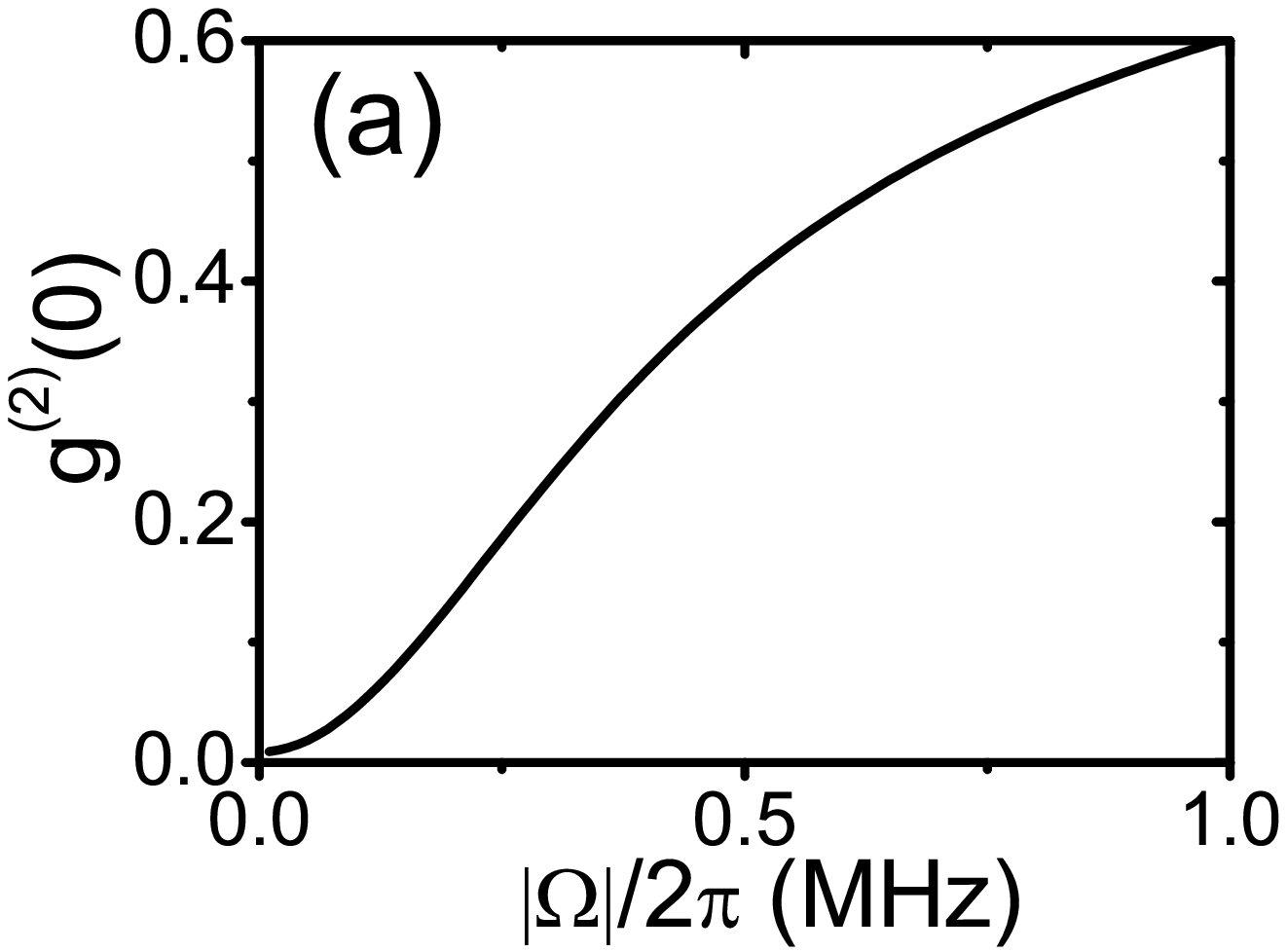}\\
\epsfxsize=8cm\epsfbox{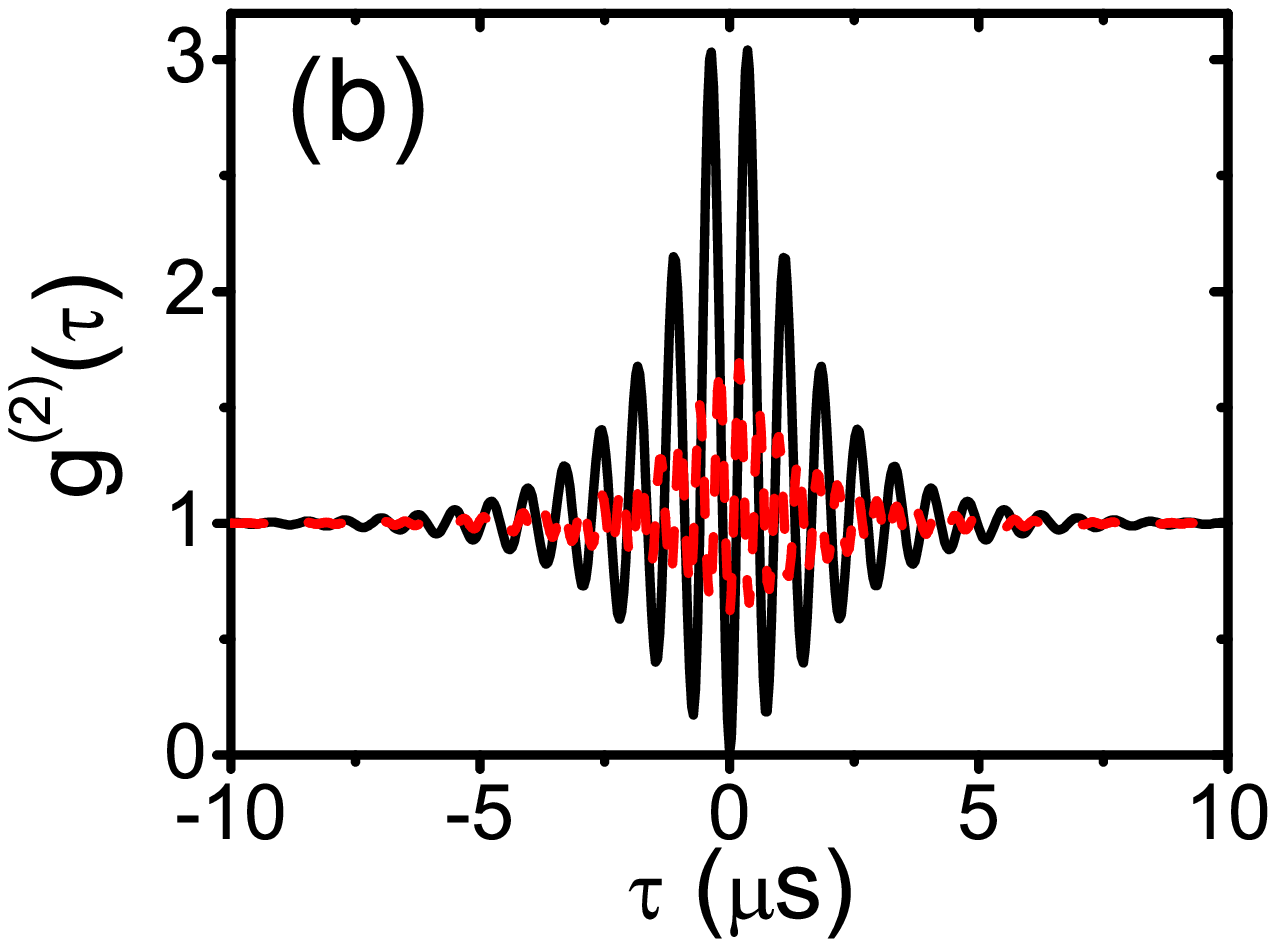}\\
\epsfxsize=8cm\epsfbox{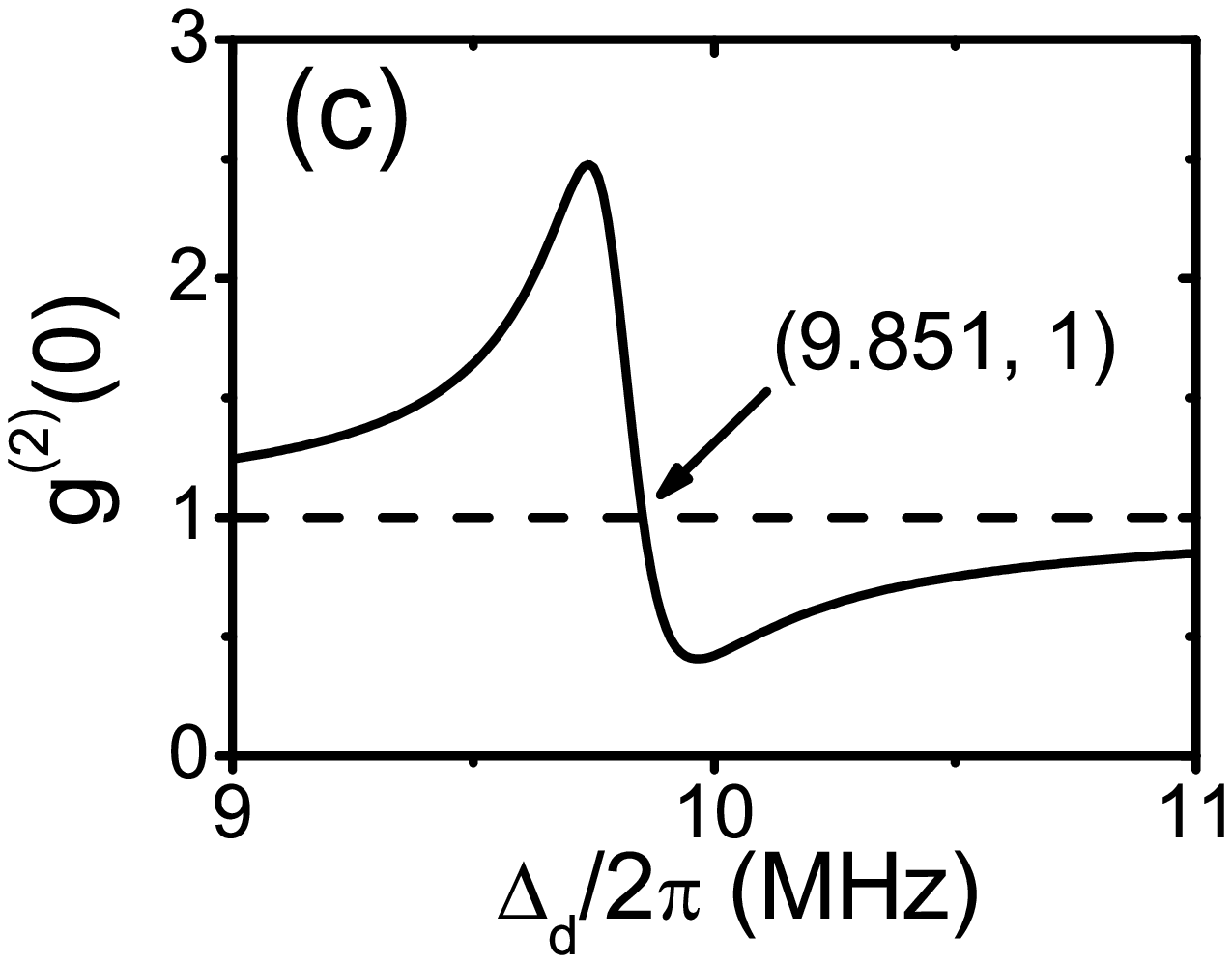}\\
\epsfxsize=8cm\epsfbox{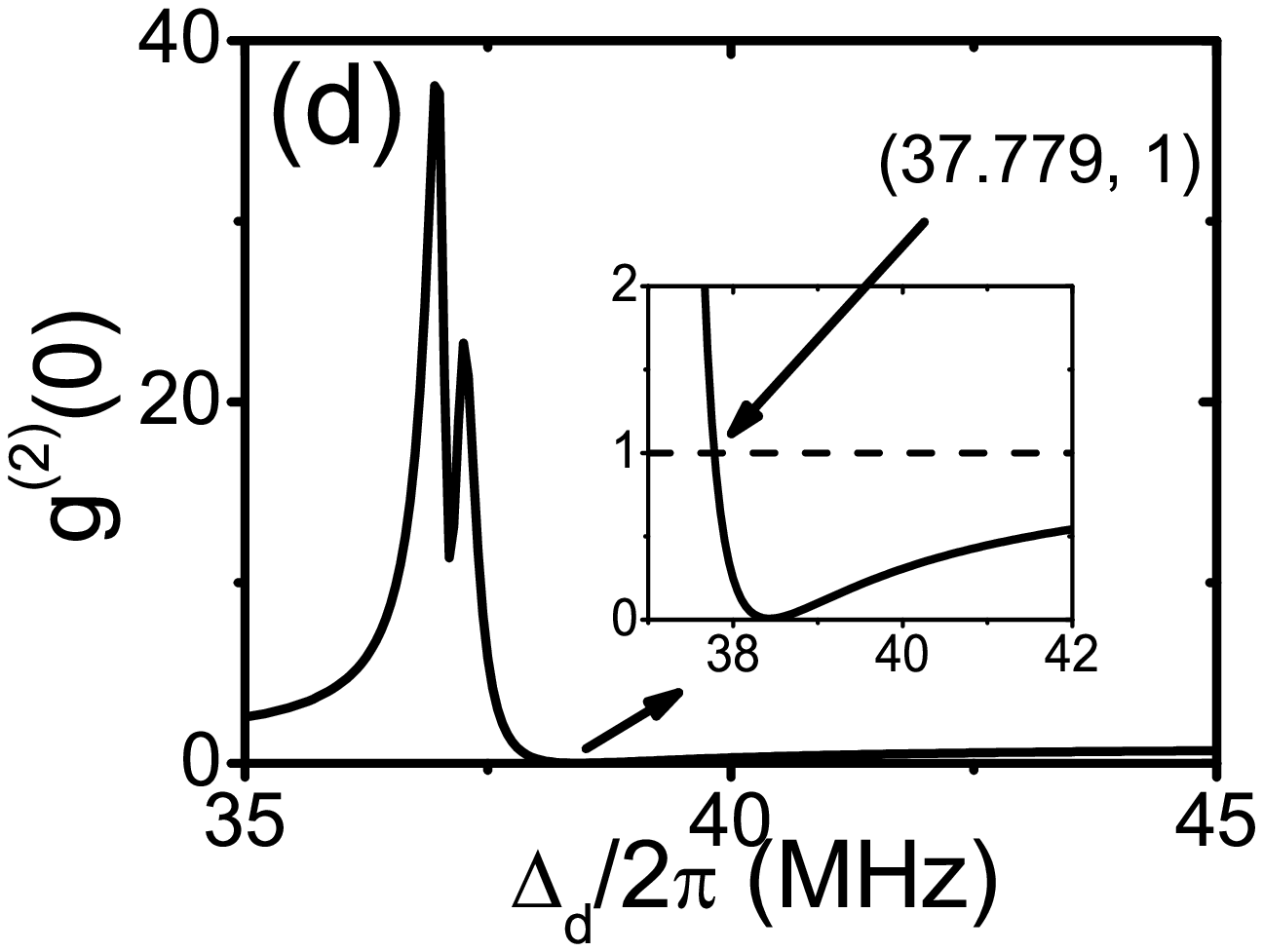} \caption[]{(Color online) (a)
Coherence $g^{(2)}(0)$ versus the driving field strength
$|\Omega|$. (b) $g^{(2)}(\tau)$ versus the delay time $\tau$,
where the solid-black and dashed-red curves correspond to the
driving strengths $|\Omega|/2\pi=0.01$  MHz and 1 MHz, respectively. Also,
$\Delta_{d}=38.5$ MHz, $\Delta/2\pi=1$ GHz, $g/2\pi=200$ MHz, and
$\kappa/2\pi=0.1$ MHz. Also, $g^{(2)}(0)$ versus the detuning
$\Delta_{d}$ in: (c) for $g/2\pi=100$ MHz and (d) for $g/2\pi=200$
MHz; with $\Delta/2\pi=1$ GHz, $|\Omega|/2\pi=0.01$ MHz, and
$\kappa/2\pi=0.1$ MHz. For parameters given in (c) and (d), the
condition of stable states for the driving field is $18\, {\rm
kHz} < \Delta_{\rm d}< 9.627\, {\rm MHz}$ and $9.1 \,{\rm kHz} <
\Delta_{\rm d}< 36.95\, {\rm MHz}$, respectively. }\label{fig3}
\end{figure}

\section{Dependence of photon blockade on detuning $\Delta_{d}$
and driving strength $\Omega$}

To show the effect of the driving field on the photon blockade,
let us now study the statistical properties of the cavity field
when the input $a_{\rm in}$  of the vacuum fields in
Eq.~(\ref{eq:3}) is considered.  We assume that the vacuum fields
$a_{\rm in}$ in Eq.~(\ref{eq:3}) result in a small fluctuation
$A_{f}(t)$ of the cavity field near its stable steady-state
$A_{s}$ by writing the cavity operator as
\begin{equation}
A(t)=A_{s}+ A_{f}(t).
\end{equation}
In addition to the steady-state solution as in Eq.~(\ref{eq:4}),
with the input $a_{\rm in}(t)=A_{\rm in}(t)e^{-i\omega_{d} t}$, we
can obtain an equation of motion for the fluctuation operator
$A_{f}(t)$ as
\begin{equation}\label{eq:5-1}
\frac{\partial A_{f}(t)}{\partial
t}=-\big[i\widetilde{\Delta}_{d}(\overline{N})+\kappa\big] A_{f}
(t)-i\alpha(\overline{N})A^{\dagger}_{f}(t)-\sqrt{2\kappa}A_{\rm
in}(t)
\end{equation}
with
\begin{eqnarray}
&&\widetilde{\Delta}_{d}(\overline{N})=\Delta_{d}-\delta\omega(\overline{N}),\\
&&\alpha
(\overline{N})=2\frac{g^4A^2_{s}\sigma_{z}}{E^{3}(\overline{N})},
\end{eqnarray}
and
\begin{equation}
\delta\omega(\overline{N})=\frac{g^2\sigma_{z}}{E^{3}(\overline{N})}
\left[E^2(\overline{N})-2g^2|A_{s}|^{2}\right].
\end{equation}
Here the terms with higher orders of $A_{f}(t)$ and
$A_{f}^{\dagger}(t)$, e.g., the term $A_{f}(t) A_{f}(t)$, have
been neglected. By applying the Fourier transform
\begin{equation}
A_{f}(t)=\int_{-\infty}^{\infty}\frac{d\omega}{\sqrt{2\pi}}\exp(-i\omega
t) A_{f}(\omega)
\end{equation}
and also using the conjugate of Eq.~(\ref{eq:5-1}) with $
A^{\dagger}_{f}(\omega)\equiv [A_{f}(-\omega)]^{\dagger}$, we
obtain the solution of the fluctuation as
\begin{eqnarray}\label{eq:6-1}
A_{f}(\omega)=\frac{i\sqrt{2\kappa}}{d(\omega)}\left[\left(\widetilde{\Delta}_{\rm
d}-\omega+i\kappa\right)A_{\rm
in}(\omega)+\alpha(\overline{N})A^{\dagger}_{\rm
in}(\omega)\right]
\end{eqnarray}
with the denominator factor
\begin{equation}
d(\omega)=(i\omega+\kappa)^2+\widetilde{\Delta}^2_{d}-|\alpha(\overline{N})|^2.
\end{equation}
The statistical properties of the cavity field can be described
via the second-order degree of coherence $g^{(2)}(\tau)$. Because
we assume that the vacuum input is Gaussian and satisfies the
Markov correlation, thus $g^{(2)}(\tau)$ can be obtained by just
calculating the correlation function with two operators using
Wick's theorem, that is
\begin{eqnarray}\label{eq:7}
g^{(2)}(\tau)&=&\frac{2{\rm Re} [A_{s}^{2}\langle
A_{f}^{\dagger}(t) A^{\dagger}_{f}(t^{\prime})\rangle+
|A_{s}|^2\langle
A^{\dagger}_{f}(t) A_{f}(t^{\prime})\rangle]}{(|A_{s}|^2+\langle n_{f}(t)\rangle)^2}\nonumber\\
&+&\frac{|\langle  A_{f}^{\dagger}(t) A^{\dagger}_{f}(t')\rangle
|^2+ |\langle A^{\dagger}_{f}(t^{\prime})
A_{f}(t)\rangle|^2}{(|A_{s}|^2+\langle n_{f}(t)\rangle)^2}+1
\end{eqnarray}
with $t^{\prime}=t+\tau$ and $\langle n_{f}(t)\rangle= \langle
A^{\dagger}_{f}(t) A_{f}(t)\rangle$.

It is easy to obtain $g^{(2)}(\tau)$ straightforwardly  by
calculating all correlation functions in Eq.~(\ref{eq:7}) using
Eq.~(\ref{eq:6-1}) and $[A_{f}(-\omega)]^\dagger$. The
second-order degrees of coherence $g^{(2)}(0)$ and $g^{(2)}(\tau)$
are plotted in Figs.~\ref{fig3}(a,b) using
experimentally-accessible parameters, e.g., in
Ref.~\cite{Hoffman11}. Figures~\ref{fig3}(a,b) show that the
cavity field tends to the classical behavior when increasing the
strength $\Omega$. In this case, the photons might not be
blockaded and can transparently pass through the circuit QED
system. To explore the effect of the frequency of the driving
field, $g^{(2)}(0)$ is plotted as a function of the detuning
$\Delta_{d}$ in Figs.~\ref{fig3}(c,d) for different coupling
strengths between the qubit and the cavity field with other
parameters given in the caption of Fig.~\ref{fig3}.
Figures~\ref{fig3}(c,d) clearly demonstrate that the nonclassical
behavior of the cavity field is out of the bistable regime for
detuning $\Delta_{d}$. For example, Fig.~\ref{fig3}(c) shows
$g^{(2)}(0)\leq 1$ when $\Delta_{d}/2\pi\geq 9.85$ MHz, which is
larger than the upper bound value $9.627$ MHz of $\Delta_{d}$ for
the bistabilility, and thus the photon blockade cannot occur in
the bistable regime.

\section{Photon blockade and transparency}

To further discuss properties of the light field transmission when
the photons are blockaded, we now study the response of the
circuit QED system to the vacuum input field, which can be
considered as a weak detecting field. According to the
input-output theory~\cite{Walls94}, the output field can be
expressed as
\begin{equation}
a_{\rm out}(t)=\sqrt{2\kappa} a(t)+a_{\rm
in}(t)+i\frac{\Omega}{\sqrt{2\kappa}}\exp(-i\omega_{d} t)
\end{equation}
by using the cavity field and the input fields. From the former
study, we know
\begin{equation}
a(t)=A(t)\exp(-i\omega_{d}t)=\left[A_{s}+A_{f}(t)\right]\exp(-i\omega_{d}t).
\end{equation}
Thus, the Fourier components of the output field can be written as
\begin{equation}\label{eq:9}
a_{\rm out}(\omega)=A_{\rm
in}(\omega^{\prime})+\sqrt{2\kappa}A_{f}(\omega^{\prime})
+\left[\frac{2\kappa A_{s}
+i\Omega}{\sqrt{2\kappa}}\right]\delta(\omega^{\prime}),
\end{equation}
with $\omega^{\prime}=\omega+\omega_{d}$ and $A_{f}(\omega)$ given
in Eq.~(\ref{eq:6-1}). The physical meaning becomes clear if the
vacuum input $a_{\rm in}(t)$  is assumed as a single-mode field
$a_{\rm in}(t)=\xi \exp(-i\omega t)$ with the real parameter $\xi
\ll |\Omega|$; that is, this weak detecting field does not change
the statistical properties of the cavity field. In this case,
$\omega^{\prime}$ in Eq.~(\ref{eq:9}) is changed to
$\omega^{\prime}=\omega_{d}-\omega$, and the terms $A_{\rm
in}(\omega^{\prime})$ and $\delta(\omega^{\prime})$ in
Eq.~(\ref{eq:9}) denote the response of the system to the input
vacuum field and the classical driving field, respectively.
However, the term with $A^{\dagger}_{\rm in}(\omega^{\prime})$
exhibits four-wave mixing with frequency $\omega-2\omega_{d}$,
which will be studied elsewhere.

\begin{figure}
\epsfxsize=8.5cm\epsfbox{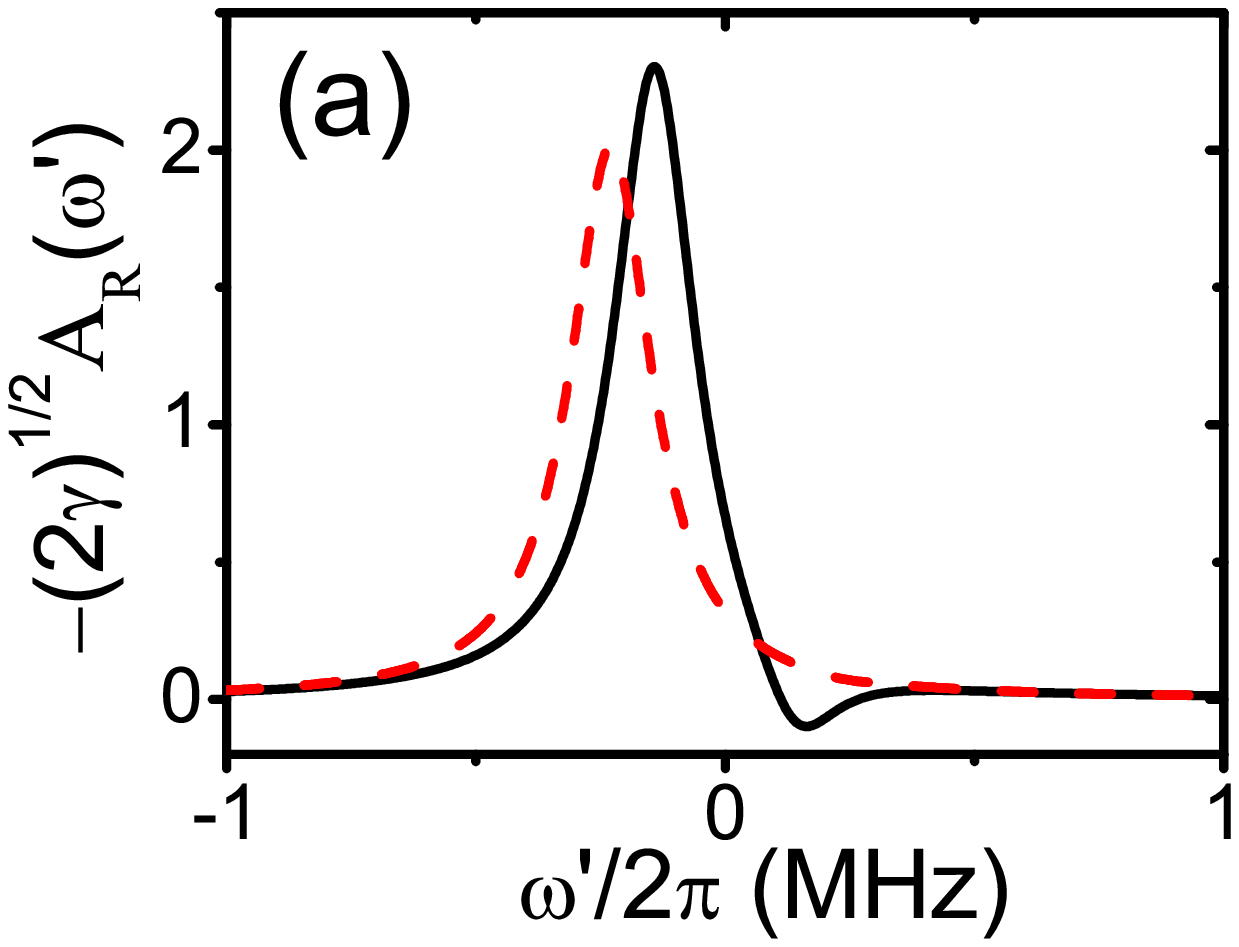}
\epsfxsize=8.5cm\epsfbox{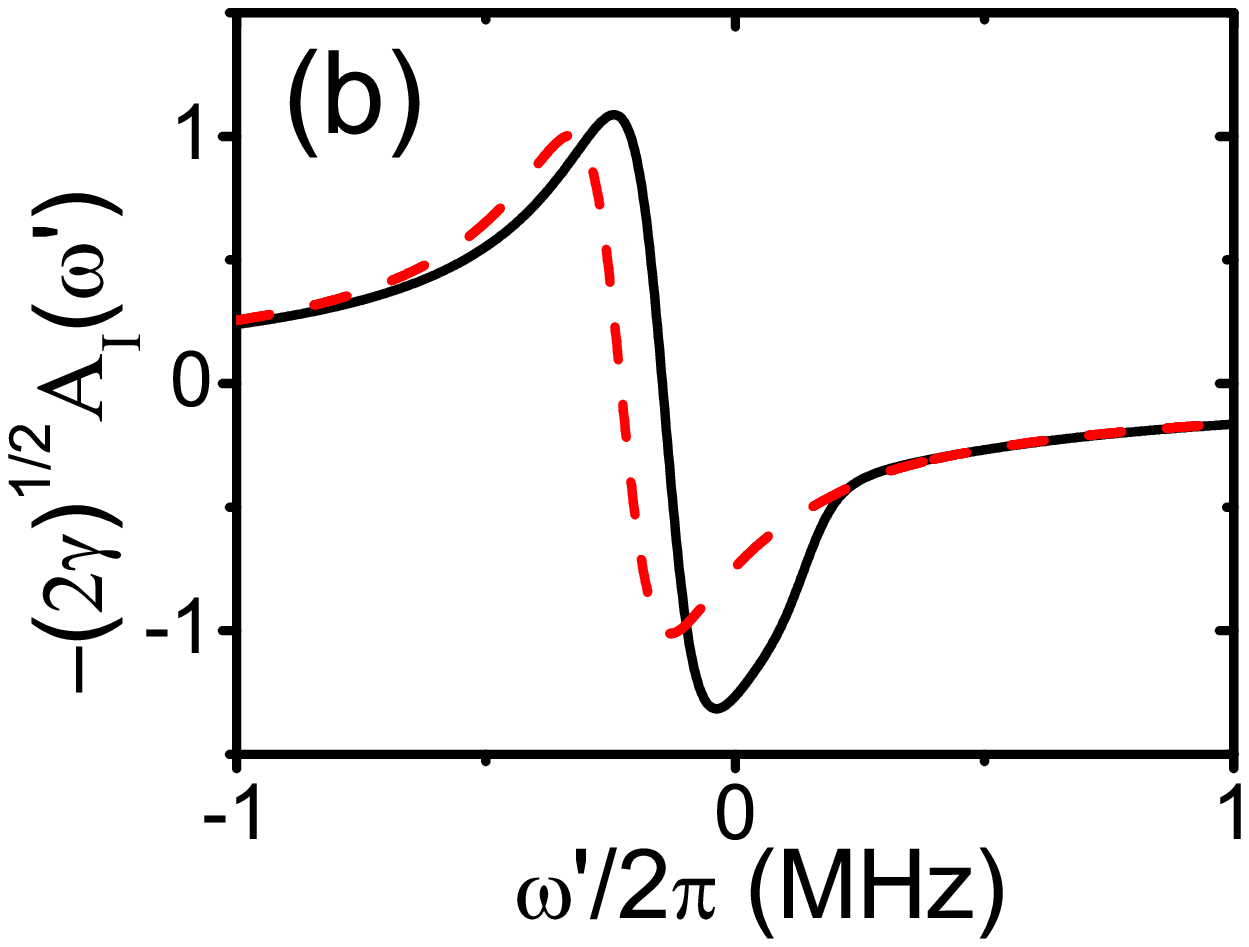}
\caption[]{(Color online) The real $A_{\rm R}(\omega^{\prime})$
and imaginary $A_{\rm I}(\omega^{\prime})$ parts $A_{\rm
in}(\omega^{\prime})/\xi$ as a function of
$\omega^{\prime}=\omega_{d}-\omega$ are plotted in (a) and (b),
respectively, for the detuning $\Delta_{d}/2\pi$ equal to 9.74 MHz
(solid black curves) and 9.96 MHz (dashed red curves) assuming
$\Delta/2\pi=1\,$GHz, $g/2\pi=100\,$MHz, and
$\kappa/2\pi=|\Omega|/2\pi=0.1\,$MHz. }\label{fig4}
\end{figure}

The coefficient of $A_{\rm in}(\omega^{\prime})$ for the output in
Eq.~(\ref{eq:9}) corresponds to that of $A_{\rm
in}(\omega^{\prime})$ in the expression $A_{f}(\omega^{\prime})$
of Eq.~(\ref{eq:6-1}) plus one. Thus, in Figs.~\ref{fig4}(a,b),
the real $A_{\rm R}(\omega^{\prime})$ and imaginary $A_{\rm
I}(\omega^{\prime})$ parts for the normalized coefficient of
$A_{\rm in}(\omega^{\prime})/\xi$ in Eq.~(\ref{eq:6-1}) for
$A_{\rm in}(t)=\xi \exp[-i(\omega-\omega_{d}) t]$ are plotted
using the same parameters as in Fig.~\ref{fig3}(c), with
$\Delta_{d}\approx 9.96$ MHz and $\Delta_{d}\approx 9.74$ MHz.
These two values correspond to the minimum (photon blockade) and
the maximum (classical case) of $g^{(2)}(0)$ in
Fig.~\ref{fig3}(c). As expected, we find that the response of the
circuit QED system to the input field (or, say, weak-detecting
field) has a Lorentzian shape, which is the same as the decay
spectrum of the single photon, when the photon is blockaded.
However, the weak detecting field shows transparency windows to
the circuit QED system when the driving field is changed such that
the cavity field is in the classical regime (or the photon is not
blockaded). Thus, we can control the photon (from blockade to
transparency) by changing the applied classical field.

\section{Conclusions}

In summary, we have studied the tunable transmission from the
photon blockade to the photon transparency in superconducting
circuit QED systems when the interaction between the qubit and the
cavity field is in the dispersive regime. We analyze the effect of
the driving field on the photon blockade. We also show the
relation between the optical bistability and the photon blockade.
We find that the photon blockade can be controlled by a classical
driving field, that is, the photon blockade strongly depends on
the properties of the driving field. We also find that the circuit
QED system can be used to generate a four-wave mixing signal. All
parameters in our numerical calculations are taken from
experimentally available data. Therefore, our study should be
experimentally realizable with current technology.  We finally
point out that the similarity between the photon blockade and
Coulomb blockade (or superconducting charge qubit) makes it
possible to simulate the electron behavior (or Josephson effect)
using photonic devices.

\section{Acknowledgement}
Y.X.L. is supported by the National Natural Science Foundation of China
under Grant Nos. 61025022, 91321208, and the National Basic Research Program
of China Grant No. 2014CB921401. A.M. is supported by Grant No. DEC-2011/03/B/ST2/01903
of the Polish National Science Centre. F.N. is partially supported
by the RIKEN iTHES Project, MURI Center for Dynamic
Magneto-Optics, JSPS-RFBR contract No. 12-02-92100, Grant-in-Aid
for Scientific Research (S), MEXT Kakenhi on Quantum Cybernetics,
and the JSPS via its FIRST program.

\end{document}